# QUANTIFYING SMOOTH MUSCLES REGIONAL ORGANIZATION IN THE RAT BLADDER USING IMMUNOHISTOCHEMISTRY, MULTIPHOTON MICROSCOPY AND MACHINE LEARNING

A.Asadbeygi[1], Y.Tobe[1], N.Yoshimura[2], S.D.Stocker[3], S.Watkins[4], A.M.Robertson[1]

[1] Department of Mechanical Engineering and Materials Sciences, University of Pittsburgh, Pittsburgh, Pennsylvania, USA, ala320@pitt.edu, yat28@pitt.edu, rbertson@pitt.edu
[2] Department of Urology, University of Pittsburgh, Pittsburgh, Pennsylvania, USA, nyos@pitt.edu@pitt.edu
[3] Department of Neurobiology, University of Pittsburgh, Pittsburgh, Pennsylvania, USA, stockers@pitt.edu
[4] Center for Biologic Imaging, University of Pittsburgh, Pittsburgh, Pennsylvania, USA, simon.watkins@pitt.edu

## SUMMARY

The smooth muscle bundles (SMB) in the bladder act as contractile elements which enable the bladder to void effectively. In contrast to skeletal muscles, these bundles are not highly aligned, rather they are oriented more heterogeneously throughout the bladder wall. In this work, for the first time, this regional orientation of the SMBs is quantified across the whole bladder, without the need for optical clearing or cryosectioning. This information is essential for biomechanical models of the bladder that include contractile elements.

## 1 INTRODUCTION

The urinary bladder's ability for periodic voiding is predominantly governed by the organized interaction between its neuronal inputs and the muscular layers comprising its wall. Central to this are the smooth muscle cells (SMCs) and bundles, collectively forming the detrusor muscle layer, which is responsible for the contraction and relaxation phases [1, 2]. The role of smooth muscle cells and bundles is paramount in the bladder's active function. These muscle cells respond to neuronal signals by contracting or relaxing, thereby changing the bladder's volume and pressure [1, 2].

Moreover, in pathophysiological conditions such as bladder outlet obstruction (BOO), the organization and functionality of these SMCs can undergo significant remodeling in both quantity and organization. This can lead to alterations in bladder compliance, hypertrophy and ultimately, detrusor overactivity or underactivity. Such changes not only play a pivotal role in causing clinical symptoms but also contribute significantly to the progression of bladder disease. Quantifying the orientation of bladder SMBs is therefore essential for developing high fidelity and anatomically accurate digital twins of the bladder to study disease and design patient specific treatments. Such digital twins, which represent a significant advancement in computational modeling, rely on accurate representation of the anisotropic and heterogeneous organization of the SMBs (contractile elements) to accurately simulate individual bladder dynamics.

Previous studies have qualitatively demonstrated the multidirectional orientation of bladder smooth muscles using electron microscopy [3] or 2D histological scans [4, 5]. Although the overall directionality of the SMC bundles has been quantified in 2D histological sections [5] (which has limitations in quantifying out of plane bundles), no study has quantitatively analyzed the 3D organization of smooth muscle bundles throughout the whole mount of a healthy bladder.

In this work, for the first time the regional orientation of smooth muscle bundles in the whole rat bladder is quantified. A new technique for 3D imaging and quantification of SMC orientation is utilized which eliminates the need for optical clearing or cryosectioning of the tissue and is illustrated on a rat bladder.

## 2 METHODOLOGY

### 2.1 Tissue harvesting and bladder inflation
A 10-week-old Sprague-Dawley male rat bladder was harvested and catheterized through the urethra. A syringe needle was inserted into the catheter tube and water-sealed with super glue. A one-way valve was attached to the needle head, to prevent voiding. A syringe was attached to the valve and 1 ml of physiological buffer solution (PBS) was injected to inflate the bladder. The final volume of bladder was measured as 1.24 ml, estimated from dimensions of major and minor axes (Figure1a).

### 2.2 Tissue fixation and immunohistochemistry
The inflated bladder attached to the one-way valve was immersed in 4% paraformaldehyde (PFA) for 12 hours for fixation. The bladder was discharged and was cut in half in the longitudinal direction, having one ureter at each side (Figure 1b, c). To stain for smooth muscle bundles, first the two halves were immersed in blocking reagent (4% normal horse serum) for six hours. The fixed sample was stained with monoclonal mouse anti-human smooth muscle actin clone 1A4 IgG2a antibody (Dako, Denmark) as the primary antibody and Alexa Fluor 568 goat anti-mouse IgG2a (Invitrogen, USA) for the secondary antibody.

### 2.3 Tissue mounting and multiphoton scanning
In order to image the anatomical organization of the SMCs, the ellipsoidal shape of the bladder should be preserved. For that purpose, a support with similar curvature to the bladder lumen was designed and 3D printed along with a sample holder. Each half of the bladder was mounted on the support, placed in the sample container and imaged using a scanning Nikon Multiphoton Microscope (MPM), (Nikon A1R MP HD, Tokyo, Japan). All the scans were performed with 10X lens at 1.24 um/pixel resolution. For each half, first a 3D MPM scan of the entire half bladder was performed, from the abluminal side. Due to the curvature of the sample, the sample edges were not captured in the whole scan; so regional MPM scans of the eight labeled zones were also performed, resulting in 5 scans per bladder half, Figures 1b and 1c. For these regional scans, the bladder was rotated so the region of interest (ROI) was located just above the dome of the support, enabling scanning of the plane tangent to the abluminal surface of the ROI, including the regions at the edge of the specimen.

### 2.4 Image processing and orientation quantification
Stacks of 3D MPM slices were projected onto a 2D plane using IMARIS 10.1 (Oxford Instruments plc). Regional intensity correction was applied to the projected images using Fiji [6]. We used the CT-FIRE package [7] to quantify the orientation of the SMC bundles. The raw MPM images contains significant noise and a high number of intersecting fibers, which results in significant error in bundle tracing. To overcome this challenge, we utilized Labkit plugin [8] in Fiji to first train a classifier model which can segment-out the bundle structures from the noisy images. Filters including gaussian blur, gaussian gradient magnitude, laplacian of gaussian, hessian eigenvalues, structure tensor eigenvalues, gabor and sobel gradient were used to create feature stacks for each image. Finally, a random forest classifier was trained based on the feature vectors and grand truth labels. Although the traditional machine learning models like random forest have much lower computational cost compared to deep learning models, they have shown very high accuracies specifically on **small size training data** [9-11]. Once the classifier is trained on just one of the labeled projected images, it can effectively segment out the SMBs in all other MPM scans. Finally, instead of the raw images, these automatic SMB segmentations with slight manual modifications were fed to the CT-FIRE, for fiber tracing and angle measurements.
Due to the circular structure of the data (angles), a Python script was written for visualizing the data in the polar histogram format.

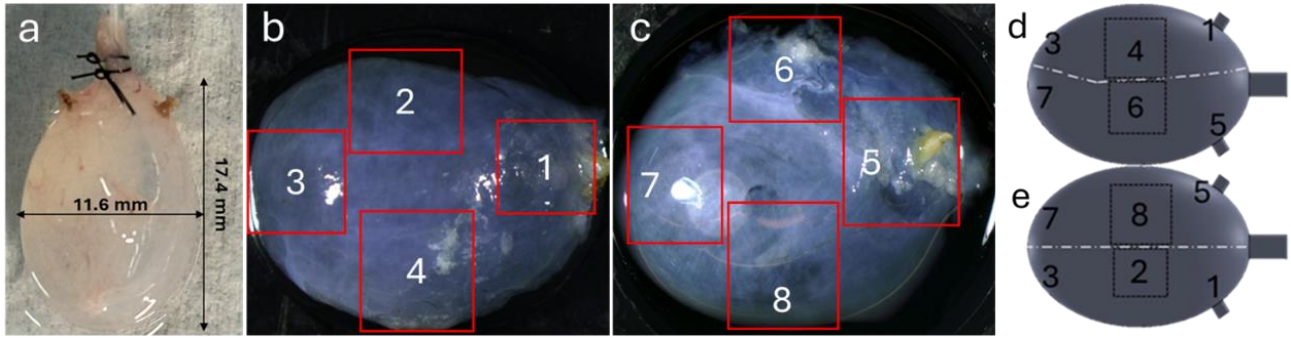

Figure 1. a) The inflated bladder. The fixed bladder bisected into (b) medial and (c) lateral halves. In b and c, samples are mounted on curved 3D-printed supports, shown with labeled zones of regional MPM imaging. Images in b,c are rotated 90° relative to a, with urethra at the top of a, and right of b,c. The relative location of each zone is shown in (d) ventral and (e) dorsal sides of the bladder.

## 3 RESULTS AND CONCLUSIONS

We first consider the projected images from the large scale MPM scans of the medial and lateral halves of the bladder, Figures 2a and 2b, respectively. Projected substacks are also shown to evaluate changes in SMB directions at different layers, Figure 2d-f. Consistent with previous studies, the direction of bundles changes through the wall thickness [12], in some but not all the regions (comparing top and bottom layers in Figure 2c-2f). Additionally, the SMC bundles display a heterogeneous directionality across the bladder (quantified in Fig. 3). Although, previous studies in the literature report longitudinal and circumferential directions for SMC bundles in pig bladder [13], our results demonstrate the bundle orientations can have substantial deviation from these two directions (Figures 2a and 2b). Moreover, strikingly, there are locations where the smooth muscle cell bundles of different angles appear to intersect (singular points), Figure 2g.

The SMB orientation is quantified in Figure 3 where the regional MPM images, corresponding bundle segmentations and polar histograms are shown. The polar histograms illustrate the angle distributions of the SMC bundles with respect to the global coordinates (labeled CD and LD). The distribution of bundle angles varies among the different scanned regions. In particular, in Zones 4 and 6, the circumferential orientation predominates among SMBs, whereas in Zones 2 and 8, a longitudinal orientation is more prevalent. These zones correspond to the central bladder area, stretching between the apex and the ureter. This pattern suggests the presence of both circumferential and longitudinal SMBs in the central bladder, implying their contraction is not uniform. This is compatible with the physical sense that this central region should be contracted in both directions, to initiate appropriate volumetric contraction, rather than being contracted in a single direction. In contrast, in the apex region (primarily Zone 7 due to the asymmetric cut), SMBs are predominantly oriented in the circumferential direction (90°). Meanwhile, Zone 3, located just below the apex, displays circumferential bundles at its junction with Zone 7 but also includes longitudinal bundles throughout the rest of the area.

Finally, Zone 5 illustrates SMBs that are predominantly oriented in the circumferential direction in the ureter zone of the lateral half, while Zone 1 in the medial half is mainly oriented in the longitudinal direction. This indicates an asymmetric orientation pattern in the ureter region between the medial and lateral halves.

It can be conjectured that this heterogeneous organization will enable a spatially non-uniform contraction pattern during voiding, in which the dome and ureteral regions squeeze circumferentially, with a bidirectional contraction in the central regions. This spatially heterogeneous deformation pattern is likely essential for efficient contraction of the bladder, needed for healthy voiding function.

**Acknowledgement** : The authors gratefully acknowledge funding from **NIH R01 DK133434** and **R01 1S10OD025041.**

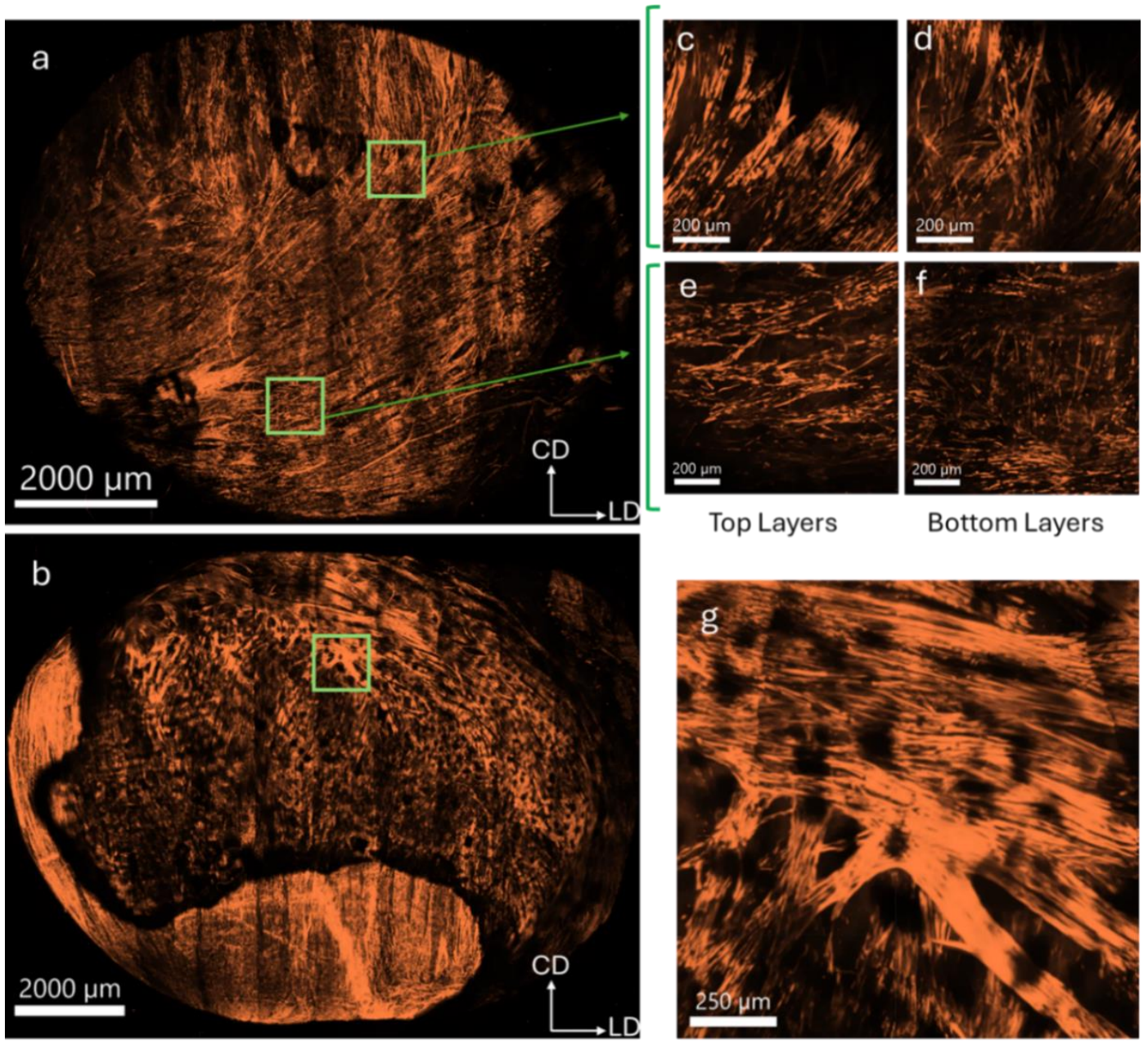

Figure 2. a and b) MPM scan of the entire medial and lateral halves, respectively. c-f) Magnified MPM images, indicating the changes in bundle direction through the wall thickness. g) Magnified image of green square in b, illustrating the intersection of bundles with different directions (singular point). LD and CD represent longitudinal and circumferential directions, respectively. In a and b, apex and urethra are located is at left and right sides, respectively.

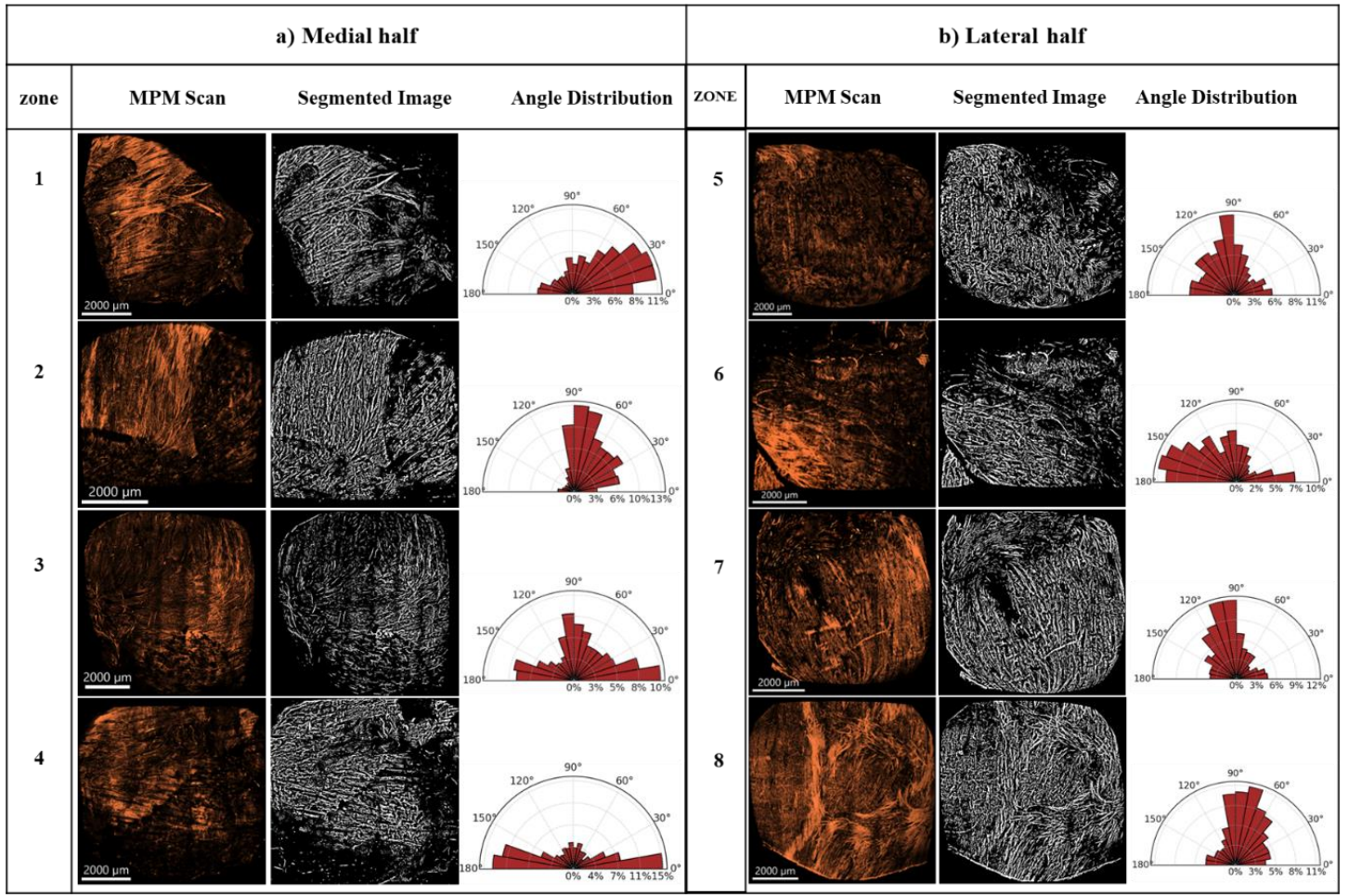

Figure 3. Polar histograms of the SMB angles, for a) medial and b) lateral halves of the bladder at 8 zones in Figures 1b and 1c.